\documentclass[journal=jpclcd,manuscript=letter,layout=onecolumn]{achemso}
\setkeys{acs}{usetitle = true}
\usepackage{booktabs} 
\usepackage{txfonts}

\title{Memory-assisted exciton diffusion in the chlorosome light-harvesting antenna of green sulfur bacteria}

\author{Takatoshi Fujita}
\affiliation{Department of Chemistry and Chemical Biology, Harvard University, Cambridge, Massachusetts 02138, USA}
\email{tfujita@fas.harvard.edu}
\author{Jennifer C. Brookes}
\affiliation{Department of Chemistry and Chemical Biology, Harvard University, Cambridge, Massachusetts 02138, USA}
\alsoaffiliation{Department of Physics and Astronomy, University College London, Gower Street, London WC1E 6BT}
\author{Semion K. Saikin}
\affiliation{Department of Chemistry and Chemical Biology, Harvard University, Cambridge, Massachusetts 02138, USA}
\author{Al\'{a}n Aspuru-Guzik}
\affiliation{Department of Chemistry and Chemical Biology, Harvard University, Cambridge, Massachusetts 02138, USA}
\email{aspuru@chemistry.harvard.edu}


\begin{document}

\begin{abstract}
Chlorosomes are likely the largest and most efficient natural light-harvesting photosynthetic antenna systems.
They are composed of large numbers of bacteriochlorophylls organized into supramolecular aggregates.
We explore the microscopic origin of the fast excitation energy transfer in the chlorosome using the recently-resolved structure and atomistic-detail simulations.
Despite the dynamical disorder effects on the electronic transitions of the bacteriochlorophylls, our simulations show that the exciton delocalizes over the entire aggregate in about 200 fs. 
The memory effects associated to the dynamical disorder assist the exciton diffusion through the aggregates and enhance the diffusion coefficients as a factor of two as compared to the model without memory.
Furthermore, exciton diffusion in the chlorosome is found to be highly anisotropic with the preferential transfer towards the baseplate, which is the next functional element in the photosynthetic system.
\end{abstract}

\begin{figure}[h]
\begin{center}
	\includegraphics[width=5cm]{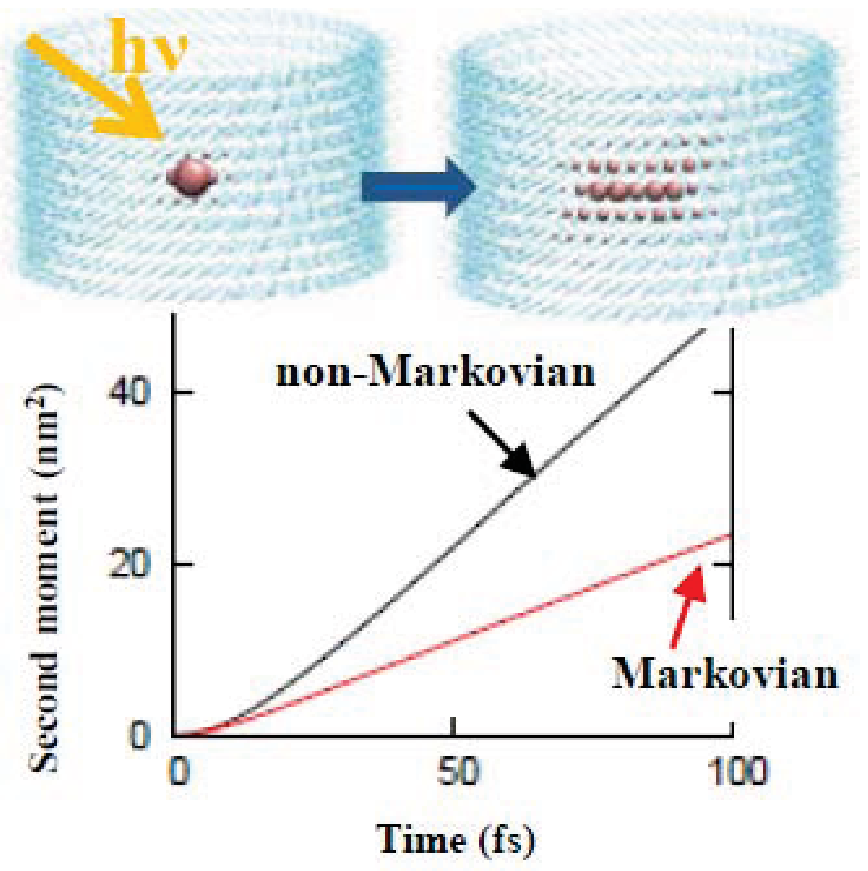}
\end{center}
\end{figure}

Chlorosomes are the largest light-harvesting antenna systems found in nature and contain hundreds of thousands of bacteriochlorophylls (BChls) enclosed by a lipid monolayer~\cite{Olson1998,Blankenship2004,Oostergetel2010,Marriott2011}.
They can capture light and transfer it in a form of molecular excitations -- excitons -- to a reaction center with high efficiency.
This feature allows bacteria to survive at very low light intensities of around 4 $\mu$mol~photons~m$^{-2}$~s$^{-1}$~\cite{Overmann1992,Beatty2005}.
Electron microscopy (EM)~\cite{Staehelin1978,Oostergetel2007} studies have suggested that the BChl $c$, $d$, or $e$ molecules are organized into rod-like aggregates -- rolls.
Light energy absorbed by the rolls is transferred via a baseplate~\cite{Pedersen2010} to the Fenna-Matthews-Olson (FMO) complex, and finally to the reaction center, where exciton dissociation occurs.
Time scales of the excitation energy transfer (EET) from the rolls to the baseplate have been estimated by time-resolved spectroscopy~\cite{Savikhin1996,Psencik1998,Prokhorenko2000,Psencik2003,Martiskainen2009}.
The EET from BChl $c$ to baseplate BChl $a$ in chlorosome from $Chloroflexus$ $aurantiacus$ occurs for instance within 10 ps~\cite{Savikhin1996,Martiskainen2009}.
A recent study~\cite{Martiskainen2012} has suggested that intra-chlorosomal EET takes place in the time range of 117-270 fs.
It has been difficult to obtain detailed structural information of chlorosomes owing to their large structural disorder. 
Thus the microscopic origin of the fast and efficient EET is still unknown.
Recent interest in the EET in the FMO complex~\cite{Engel2007,Panitchayangkoon2010,Moheseni2008,Rebentrost2009,Rebentrost2011,Shim2012}, red algae~\cite{Collini2010,Hossein-Nejad2011}, and plants~\cite{Schlau-Cohen2012} motivates this study and suggests that a theoretical analysis of exciton dynamics in the chlorosomes might provide insights into their function as arguably nature's most efficient light-harvesting systems.

Recent understanding of biosynthesis of the BChl $c$ in $Chlorobaculum$ $tepidum$ ($C.$ $tepidum$)~\cite{Chew2007a} has led to a well-characterized $C.$ $tepidum$ three-point mutant~\cite{Chew2007b}.
This mutant synthesizes a chlorosome that contains >95\% BChl $d$ and forms regularly-packed and tubular-shaped elements~\cite{Oostergetel2007}.
Ganapathy \textit{et al}.,~\cite{Ganapathy2009} have proposed an atomistic structure of the rolls from the three-point mutant by using solid-state NMR and cryo-EM images:
$Syn$-$anti$ monomer stacks are basic building blocks, and they are organized into coaxial cylinders to form the tubular-shaped elements.
Recent measurement of the linear dichroism spectra of individual chlorosomes~\cite{Furumaki2011} has confirmed that the chlorosomes from the three-point mutant are less disordered than those from a wild-type, which is compatible with the highly-symmetric structure proposed by Ganapathy \textit{et al}.~\cite{Ganapathy2009}.
The atomistic structures of the roll are shown in Figure 1.
The $syn$ monomer and $anti$ monomer stacks, Figure 1 (e), run perpendicular to the roll symmetry axis; $n$-BChl pairs form a ring with $n$-fold rotational symmetry.
These rings are stacked in-parallel so that they constitute a helical cylinder (Figure 1(a)).
Several concentric cylinders are further combined with a spacing of the order of 2 nm (Figure 1(d)).

In this work, we combine molecular dynamics (MD) simulations, time-dependent density functional theory (TDDFT) excited-state calculations, and quantum stochastic propagation of the exciton dynamics to characterize exciton diffusion in the system proposed by Ganapathy \textit{et al}.~\cite{Ganapathy2009} We find that the supramolecular structure of the rolls enables transfer in the circumference direction, which is relevant for transfer to the baseplate.
The environment induces the fluctuations in the electronic transitions of the BChl in the roll.
The memory effects associated to these fluctuations assist the exciton diffusion through the roll and can result in the enhancement of the exciton diffusion as a factor of two as compared to the model without memory (Markovian model).

We briefly summarize the theoretical methods for the exciton dynamics.
Details of the calculations and underlying assumptions of the theory will be presented elsewhere.
Each Bchl $d$ is treated as a two-level system of ground and $Q_y$ excited states. 
The time-dependent excitonic Hamiltonian coupled to the environmental degrees of freedom is given by
\begin{equation}
H(t)=\sum_{m}^{N} \{ \left< \epsilon_m \right > + \Delta \epsilon_m(\mathbf{R}(t)) \}\bigl|m \rangle \langle m\bigr|  + 
\sum_{m \neq n}^N V_{mn} \bigl|m \rangle \langle n\bigr|,
\end{equation}
where the $\bigl|m\rangle$ denotes a state where excitation is localized at the $m$-th BChl (site) and all other pigments are in the ground state. $\mathbf{R}(t)$ is a nuclear configuration at time $t$.
Here, $\Delta \epsilon_m(\mathbf{R}(t))=\epsilon_m(\mathbf{R}(t))-\left< \epsilon_m\right >$, $\epsilon_m(\mathbf{R}(t))$ is an excitation energy of $m$-th site in the nuclear configuration $\mathbf{R}(t)$, and the bracket $\left<  \right>$ denotes an average over the nuclear configurations. 
The electronic coupling between $m$-th and $n$-th sites $V_{mn}$ is calculated within the point-dipole approximation.
It is assumed that transition dipole moments are oriented along the line connecting $N_{\rm{I}}$ to $N_{\rm{III}}$ atom in a BChl and have 30 Debye$^2$ as the squared dipole strength estimated from the experimental data~\cite{Prokhorenko2000}.

The time evolution of the exciton system is given by the time-dependent Schr$\rm{\ddot{o}}$dinger equation:
\begin{equation}
i\hbar \frac{\partial}{\partial t}\bigl| \psi(t) \rangle=H(t)\bigl| \psi(t) \rangle.
\end{equation}
Following our previous work on the FMO complex~\cite{Shim2012}, we take a stochastic approach to obtain the fluctuations of the site energies.
Site energy fluctuations are obtained by solving a set of Langevin equations with the white noise
\begin{equation}
\frac{\partial}{\partial t}\Delta \epsilon_m(t)=-\frac{\Delta \epsilon(t)}{\tau_m}+F_m(t),
\end{equation}
where $F_m(t)$ is a random force; $\left <F_m(t) \right>=0$, and $\left < F_m(t)F_m(0)\right>=2\left<\Delta \epsilon_m^2\right>\delta(t)/\tau_m$.
According to the eq. 3, the site energy correlation function is  $\left < \Delta \epsilon_m(t)\Delta \epsilon_m(0)\right>=\left<\Delta \epsilon_m^2 \right>\exp{(-t/\tau_m)}$.
Therefore, the exciton dynamics governed by eq. 2 is equivalent to the Kubo-Anderson (KA) stochastic model~\cite{Anderson1954,Kubo1954}.
Here, a relaxation time of the site energy fluctuation $\tau_m$ is defined as an integration of an autocorrelation function (ACF) of the site energy.
The density matrix is obtained as an ensemble average of these unitary evolutions, $\rho(t)=(1/M)\sum_i^M \bigl| \psi_i(t) \rangle \langle \psi_i(t) \bigr|$.
The input parameters to this description of the exciton dynamics are $\left< \Delta \epsilon_m\right>$, $\left<\Delta \epsilon_m^2 \right>$, and $\tau_m$, which are obtained from MD/TDDFT calculations.
We use a model of ten-stacked rings, Figure 1 (a), for the exciton dynamics in the chlorosome, where each ring consists of 60 BChls.
Due to the symmetry of the roll, we can assume that the statistical properties of the environmental noise are similar for all the $syn$-$anti$ stacked monomers.
Therefore, a random sampling of MD/TDDFT trajectories for a single dimer of $syn$ and $anti$ BChls can be used for the whole roll.

To include the contributions from several rolls, we set up a model of three-concentric five-stacked rings, Figure 1 (c) and (d), for MD/TDDFT calculations.
Rings from different layers consist of 60, 80 or 100 BChl molecules, and the total number of the BChl molecules is thus 1200.
A time step of 2 fs was employed, and the temperature was set at 300 K by the Langevin thermostat in the MD.
The hydrogen atoms were constrained at their ideal positions using the SHAKE algorithm.
After a 1 ns equilibration run, 20 ps production run was carried out and stored every 4 fs.
The MD simulations were performed by a CUDA implementation of NAMD~\cite{NAMD,Stone2007}.
Using the snapshots from the 20 ps trajectory, the $Q_y$ excitation energies for the $syn$ and the $anti$ monomers were obtained to calculate the averages and correlation functions of the site energies.
TDDFT excited-state calculations were performed with the long-range corrected hybrid functional of Becke ($\omega$B97)~\cite{Chai2008} and 6-31G basis set in the presence of external point charges of all other molecules, as implemented in the Q-CHEM quantum chemistry package~\cite{Q-Chem}.

The average site energies were calculated to be 15243.0 and 15319.4 cm$^{-1}$ for the $syn$ and the $anti$ monomers, respectively, and the corresponding standard deviations were 532.2 and 556.5 cm$^{-1}$.
Although the $syn$ and the $anti$ monomers are not in same environment, the difference in the averages and standard deviations of the site energy are small.
An absorption maximum was estimated by diagonalizing the Hamiltonian without $\Delta \epsilon_m(t)$ and was obtained as 13220.6 cm$^{-1}$.
This value is in reasonable agreement with the experimental absorption maximum of 13888.9 cm$^{-1}$~\cite{Furumaki2011}. 
The ACFs and spectral densities for the $syn$ and the $anti$ monomers are shown in Figure 2.
The spectral densities were calculated as a weighted cosine transform of the correlation function~\cite{Damjanovic2002,Olbrich2011b}:
\begin{equation}
J_m(\omega)=\frac{2}{\pi\hbar}\tanh{\left( \frac{\hbar\omega}{2k_BT} \right)}\int_0^{\infty}dt \left<\Delta \epsilon_m(t)\Delta \epsilon_m(0) \right>\cos{(\omega t)}.
\end{equation}
The calculated relaxation times for the $syn$ and the $anti$ monomers are 8.80 and 4.51 fs, respectively.
It is remained to be elucidated whether this difference has some physical importance or it stems from the approximations of the model.
The spectral densities show the vibrational modes in the roll coupled to the excitation energies.
Strong peaks around 1600 to 2000 cm$^{-1}$ can be attributed to internal bond stretching in the porphyrin macrocycle of the BChl.
Characteristic peaks are found in the low-frequency region, which are ascribed to intermolecular interactions arising from the closed-packed structure.
Reorganization energies were calculated by $E_{R}=\int^{\infty}_{0}d\omega J(\omega)/\omega $ and are 266.5 and 298.4 cm$^{-1}$ for $syn$ and $anti$ monomers, respectively.

We then run exciton dynamics simulations by solving eqs. 2 and 3.
Initial conditions of the wavefunctions were chosen as a localized state on an $anti$ monomer (Figure 1 (b)), and $\Delta \epsilon_m(0)$ were sampled from the Gaussian distribution.
The density matrix was calculated by averaging 1000 realizations of the stochastic processes.
Figure 3 shows population dynamics obtained by the exciton simulations.
Excitons delocalize over the entire roll at the time scale of approximately 200 fs.
The population spreads faster in the circumference than in the coaxial directions.
This anisotropic diffusion is due to the difference in the electronic coupling constants.
Table 1 shows the electronic coupling constants between the initially excited monomer and its neighboring monomers.
The electronic coupling between $syn$-$anti$ monomers are larger than those between monomers from neighboring rings.

In order to quantify the exciton diffusion, second moments of the exciton propagation in the circumference and the coaxial directions (see Figure 1 (a)) were calculated and are shown in Figure 4 (a).
The second moments initially scale quadratically and then linearly in time.
The transition from the initial ballistic regime to the diffusive regime is observed at around 20 fs, which is similar to high exciton mobility J-aggregates~\cite{Stephanie2012}.
The second moment in the circumference direction is about four times larger than in the coaxial direction.
This anisotropic diffusion may play a role in the efficient EET: as it is proposed that the symmetry axis lies parallel to the baseplate~\cite{Pedersen2010}, the faster diffusion in the transverse direction implies the efficient EET to the baseplate.
Furthermore, the present method is compared with the Haken-Strobl-Reineker (HSR)~\cite{Haken1972,Haken1973} and the classical hopping~\cite{Ern1972} models.
In the HSR and the hopping models, a pure dephasing rate for $m$-th site $\gamma_m$ was obtained as $\gamma_m=2\left<\Delta \epsilon_m^2 \right>\tau_m/\hbar$~\cite{Breuer}. The KA model gives the second moment twice as large as those by the HSR, leading to memory-assisted diffusion (MAD).
The difference between the KA and the HSR is in the memory of the bath fluctuations (KA: $\left<\Delta \epsilon_m(t)\Delta \epsilon_m(0) \right> \propto \exp{(-t/\tau_m)}$ vs HSR: $\left<\Delta \epsilon_m(t)\Delta \epsilon_m(0) \right> \propto \delta (t)$).
This comparison indicates that the memory of the bath fluctuations enhances the diffusion.
The diffusion of the excitation energy is directed by the supramolecular arrangement and enhanced the memory of the bath fluctuations, which, according to the results in this paper, might be very essential for the fast and efficient EET.
These results are complementary to our findings that non-Markovianity is also near maximal for the FMO complex~\cite{Rebentrost2009,Rebentrost2011}.

Finally, we investigate the dependence of the exciton diffusion on static disorder.
Origins of the static disorder are, for example, inner structure disorder inside the chlorosome and existence of different homologues of the BChls.
Here, the static disorder was incorporated by introducing Gaussian random shifts in the average site energies.
Diffusion coefficients in the circumference direction were calculated as a function of a standard deviation of variable static disorder parameters and are shown in Figure 4 (b).
The HSR model gives the same diffusion constants as the hopping model. Since rate constants in the hopping model were calculated with  $\left<\Delta \epsilon_m(t)\Delta \epsilon_m(0) \right> \propto \delta (t)$~\cite{Ern1972}, their difference is only in the initial ballistic propagation up to 20 fs.
The KA models gives about twice as large diffusion coefficients as those from the HSR up to the static disorder of 400 cm$^{-1}$, while the memory effects become small in the presence of the large static disorder.
In general, the static disorder broadens an absorption band, and it will be favorable for the chlorosomes to capture a wider range of light energy. On the other hand, presence of the static disorder leads to the slower diffusions. 
The MAD may be a mechanism to overcome negative effects of the static disorder inherent in the chlorosome, whiles preserving the positive effects of static disorder, both working in concert towards the most productive EET.

We have presented the theoretical study on the exciton dynamics in the chlorosome.
We have shown that the exciton diffusion is highly anisotropic with the faster direction toward the baseplate.
The memory effects of the bath fluctuations increase the exciton diffusion coefficients by two times for broad range of the static disorder.
We have proposed that the supramolecular arrangement and the memory-assisted diffusion are the microscopic origin of the fast and efficient excitation energy transfer in the chlorosome.
The present methods can be applied to more complex systems, such as a system that includes many rolls and/or a baseplate. 
Work in such a direction will provide a further microscopic insight into the chlorosomes as well as a design principle of artificial light-harvesting systems~\cite{Scholes2011}.

\acknowledgement
The authors would like to thank Prof. Huub J. M. de Groot for fruitful discussions and the donation of the $syn$-$anti$ chlorosome configuration template.
We further appreciate St\'{e}phanie Valleau for very useful discussions.
T. F. and A. A.-G. acknowledge support from the Center for Excitonics, an Energy Frontier Research Center funded by the US Department of Energy, Office of Science and Office of Basic Energy Sciences under award DE-SC0001088.
J. C. B. acknowledges support from Welcome Trust UK.
S. K. S. and A. A.-G. also acknowledge Defense Threat Reduction Agency grant HDTRA1-10-1-0046.
Further, A. A.-G. is grateful for the support from Defense Advanced Research Projects Agency grant N66001-10-1-4063, Camille and Henry Dreyfus Foundation, and Alfred P. Sloan Foundation 

\providecommand*\mcitethebibliography{\thebibliography}
\csname @ifundefined\endcsname{endmcitethebibliography}
  {\let\endmcitethebibliography\endthebibliography}{}

\clearpage

\begin{table}
\caption{The electronic coupling constants between the initially excited monomer (1) and its neighboring monomers (2-7) in the ten-stacked rings.
An index in the row refers to the monomer labeled by the same index in Figure 1 (b)}
\begin{tabular}{crrrrrr} \\ \toprule 
                                 &   2        &            3 &        4 &        5&        6   &  7   \\ \midrule
$V_{1n}$ (cm$^{-1}$)          & --695.1    &     --671.9 &  --90.4 &   154.9& --208.2   &  --102.2  \\ \bottomrule
\end{tabular} \\
\end{table}

\clearpage

\begin{figure}[h]
\begin{center}
	\includegraphics[width=12cm]{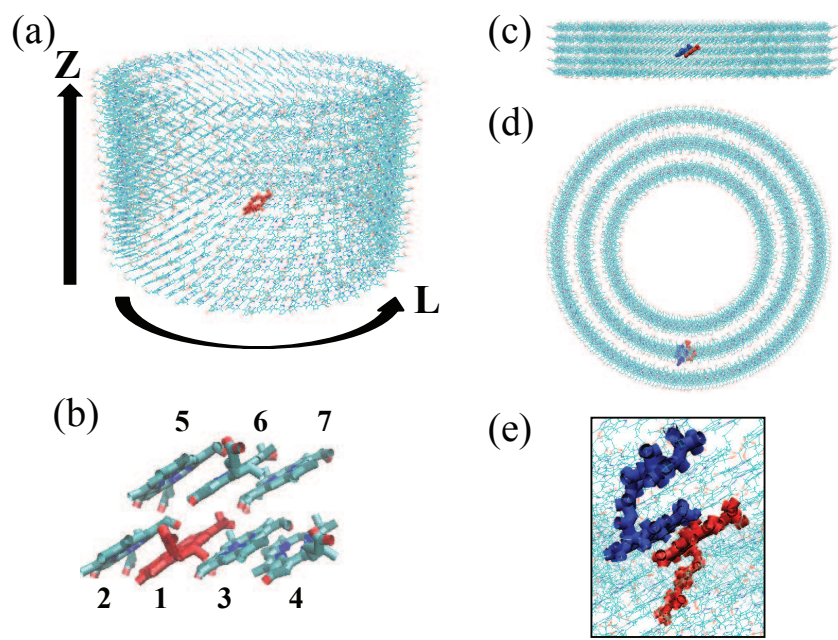}
\end{center}
\caption{(a) A model of ten-stacked rings for the exciton dynamics. (b) Initially excited monomer (red) and its neighboring monomers in the ten-stacked rings. (c) Side view and (d) top view of a model of three-concentric five-stacked rings for the MD/TDDFT calculations. (e) $Syn$ (red) and $anti$ (blue) monomers in the three-concentric five-stacked rings, the site energies of which were obtained from the TDDFT excited-state calculations.} 
\end{figure}

\clearpage

\begin{figure}[h]
\begin{center}
	\includegraphics[width=8cm]{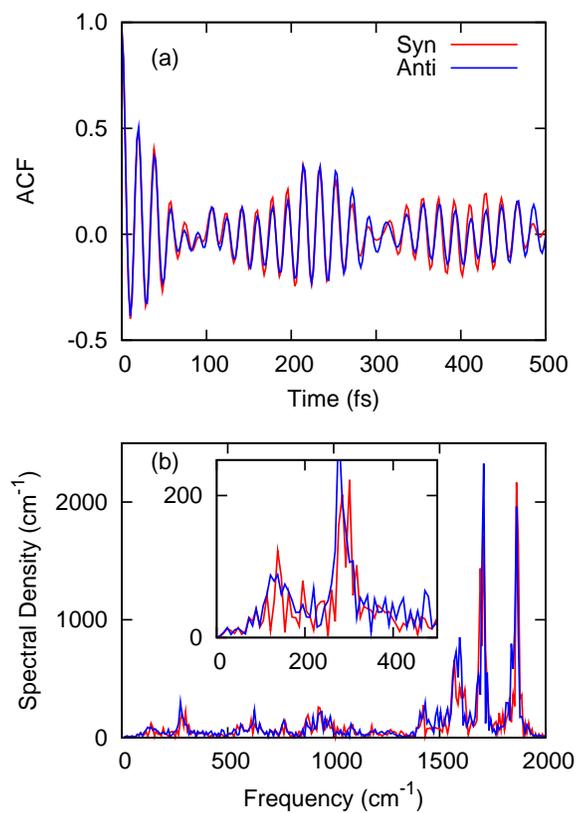}
\end{center}
\caption{(a) Autocorrelation functions (ACF) of the site energies and (b) spectral densities for the $syn$ (red) and the $anti$ (blue) monomers. Low-frequency regions of the spectral densities are shown in the inset.}
\end{figure}

\clearpage
\begin{figure}[h]
\begin{center}
	\includegraphics[width=12cm]{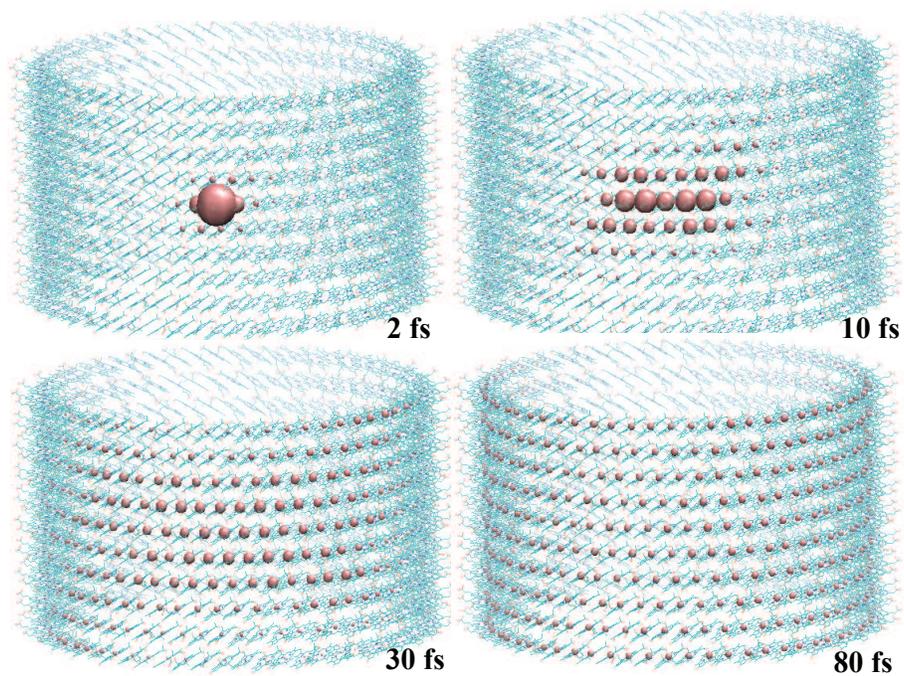}
\end{center}
\caption{Population dynamics at selected times. Pink spheres represent Mg atoms of the BChl molecules,
and the size of the Mg atom specifies the population value of the BChl molecule.}

\clearpage

\end{figure} 
\begin{figure}[h]
\begin{center}
 	\includegraphics[width=8cm]{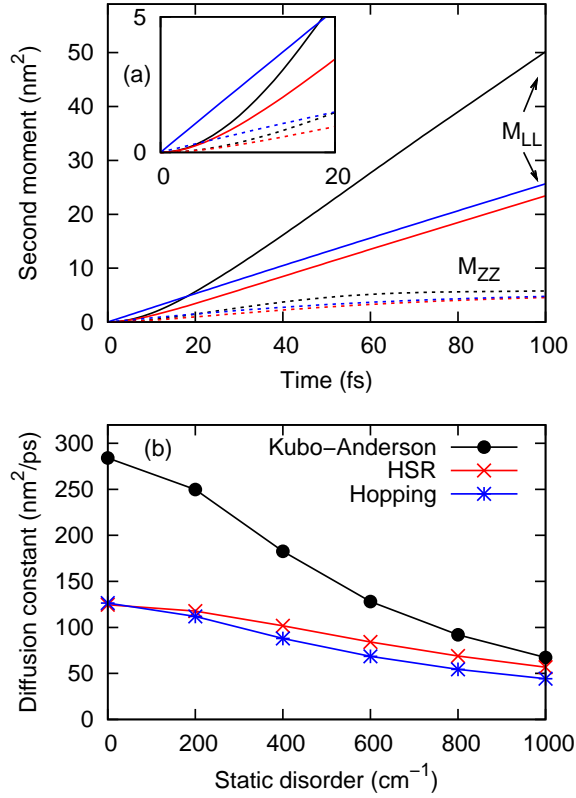}
\end{center}
\caption{(a) Second moments obtained from the Kubo-Anderson (black), the Haken-Strobl-Reineker (HSR) (red), and the classical hopping (blue) models. Solid and dotted lines refer to the second moments in the circumference ($\rm{M}_{LL}$) and coaxial ($\rm{M}_{ZZ}$) directions (see Figure 1(a)), respectively. (b) Exciton diffusion coefficients in the circumference direction as a function of a standard deviation of the static disorder.}
\end{figure}

\end{document}